\documentclass{jjap2}
\title{Extension of the simulation code ACAT to treat real atomic positions}

\author{
Arimichi \textsc{Takayama}$^{1}$\thanks{E-mail address: takayama.arimichi@nifs.ac.jp},
Seiki \textsc{Saito}$^{2}$,
Atsushi M. \textsc{Ito}$^{1}$,
Takahiro \textsc{Kenmotsu}$^{3}$,
and
Hiroaki \textsc{Nakamura}$^{1,2}$
}

\inst{$^{1}$Fundamental Physics Simulation Research Division, National Institute for Fusion Science, Toki, Gifu 509-5292 Japan.\\
$^{2}$Department of Energy Engineering and Science, Graduate School of Engineering, Nagoya University, Nagoya 464-8603 Japan.\\
$^{3}$Graduate School of Life and Medical Sciences, Doshisha University, Kyotanabe, Kyoto 610-0394 Japan.}

\abst{
We have investigated plasma-surface interactions with molecular dynamics (MD) simulations\cite{MD1,MD2}.
It, however, is high cost computation and is limited to simulations for materials of nanometer order.
In order to overcome the limitation, a complementary model based on binary collision approximation (BCA) can be established.
We employed a BCA-based simulation code ACAT\cite{ACAT} and extended to handle any structure involving crystalline and amorphous.
The extended code, named "AC$\forall$T", stores all positions of projectile and target atoms and velocities of recoil atoms, so it can be combined with the MD code.
It also holds the potential to reproduce channeling phenomena.
Thus it is expected to be useful for evaluation of channeling effects.
}

\kword{
binary collision approximation,
plasma-surface interactions,
sputtering,
graphite,
hydrogen
}

\begin{document}
\maketitle

\section{Introduction}
In nuclear fusion devices plasmas of hydrogen isotopes exist and contact material surfaces.
The "divertor configuration" is employed in order to control impurities produced by impacts of plasmas to the surface and to reduce the heat load to plasma facing materials.
In the configuration divertor plates, whose potential constituent includes carbon and tungsten, are installed.
Understanding of the divertor physics and designing an appropriate configuration is essential for the establishment of nuclear fusion reactor.
These require knowledge on plasma-surface interactions (PSI).
Thus we have investigated interactions between hydrogen atoms and carbon material, such as graphite, with molecular dynamics (MD) simulations\cite{MD1,MD2}.

The molecular dynamics simulation code solves equations of motion for all particles under the modified Brenner reactive empirical bond order (REBO) potential\cite{MD1}.
Evaluation of the REBO potential requires consideration of effects from multiple particles, and thus it is generally high cost computation.
This limits the applicable material scale length of the MD simulation to about order of nanometer and the energy range to about order of kilo electron volt.

In order to investigate PSI with numerical simulation we have to overcome the limitation contained by MD simulation.
A complementary model can be established based on binary collision approximation (BCA).
There exist intensive PSI related works with binary collision based monte carlo simulations\cite{TRIM.SP,MARLOWE,ACAT,ACOCT,DYACAT,DYACOCT}. 
The binary collision approximation simplifies interactions between material elements and reduces them to the sequence of the binary collisions.
A benefit of the model is that it is rather simple and requires less computing resources than MD model.
It, however, holds a limit of application on lower energy region.

We think that a hybrid simulation of molecular dynamics and binary collision approximation is promising.
In the integrated simulation code the part of binary collision approximation covers higher energy region, and the part of molecular dynamics governs lower energy region and solves only the vicinity of the projectile and recoil atoms.
The threshold energy between the two regions should be determined according to a validity condition of the binary collision approximation, which is roughly estimated as 200 eV.

In this paper we extend an existing binary-collision-approximation-based simulation code ACAT\cite{ACAT} in order to combine MD simulation code and BCA based one.

\section{ACAT code and its extension code AC$\forall$T}

We employ a binary-collision-approximation (BCA) based simulation code ACAT.
The ACAT code was developed to simulate \underline{a}tomic \underline{c}ollisions in an \underline{a}morphous \underline{t}arget within the framework of the binary collision approximation.
Projectile particles are traced through binary collisions.
Target particle, with which projectile collides, is randomly distributed in each unit cell whose size $R_0 = N^{-1/3}$, where $N$ is the number density of the target material.
In terms of randomly distributed target particle, this code employs the monte carlo method and aims for the atomic collisions in amorphous target.

Figure 1 depicts the trajectory of two particles interacting according to a conservative central repulsive force.
The scattering angle in the center-of-mass system (CM-system) is
\begin{equation}
\Theta = \pi - 2b\int_{r_0}^{\infty} \frac{1}{r^2g(r)} dr,
\label{eq:scattering_angle}
\end{equation}
where,
\begin{equation}
g(r) = \sqrt{1-\frac{b^2}{r^2}-\frac{V(r)}{E_r}},
\label{eq:g_r}
\end{equation}
$b$ is the impact parameter,
$E_r = E_0 m_1/(m_1+m_2)$ is the relative kinetic energy,
$E_0$ is the incident kinetic energy of the projectile,
$V(r)$ is the interatomic potential,
$r_0$ is the solution of $g(r)=0$,
$m_1$ and $m_2$ are the mass of the projectile and the target atom, respectively.

The trajectories of particles are approximated as the asymptotes of them in the laboratory system (L-system).
So they consists of linkage of straight-line segments.
The starting point of the projectile and the recoil atom after a collision is given by $\Delta x_1$  and $\Delta x_2$, which are the shifts from the initial position of the target atom shown in Fig.1:
\begin{equation}
\Delta x_1 = \frac{2\tau + (A-1) b \tan(\Theta/2)}{1+A},
\label{eq:Delta_x1}
\end{equation}

\begin{equation}
\Delta x_2 = b \tan(\Theta/2)-\Delta x_1,
\label{eq:Delta_x2}
\end{equation}
where 
\begin{equation}
\tau = \sqrt{r_0^2-b^2} - \int_{r_0}^{\infty} \left\{\frac{1}{g(r)}-\frac{r}{\sqrt{r^2-b^2}}\right\} dr,
\label{tau}
\end{equation}
and the mass ratio $A = m_2/m_1$.

As the interatomic potential $V(r)$, the Moliere approximation to the Thomas-Fermi potential \cite{Moliere} is employed:
\begin{equation}
V(r) = \frac{Z_1 Z_2 e^2}{r}\Phi(r/a),
\label{eq:Moliere_V}
\end{equation}
\begin{equation}
\Phi(x) = 0.35{\rm e}^{-0.3x}+0.55{\rm e}^{-1.2x}+0.10{\rm e}^{-6.0x},
\label{eq:Moliere_Phi}
\end{equation}
where $a$ is the screening length, and $Z_1$ and $Z_2$ are the atomic numbers of the projectile and the target atom, respectively.

The procedure of searching the collision partner is as follows:
We define a unit vector $\mbox{\boldmath $e$}_{\rm p}$ in the direction of a moving projectile and a step length $\Delta x$.
Here, the notation "projectile" means not only the incident particle but any recoil atom.
The position of the projectile in $n$-th step , which is initially $\mbox{\boldmath$R$}^{(0)}$, is $\mbox{\boldmath$R$}^{(n)} = \mbox{\boldmath$R$}^{(0)} + n\Delta x\mbox{\boldmath $e$}_{\rm p}$.
If the unit cell involving $\mbox{\boldmath$R$}^{(n)}$ is different from that of the initial position $\mbox{\boldmath$R$}^{(0)}$, a target atom is produced in the new unit cell by use of four random numbers.
Three random numbers are for location and one is for the kind of target atom.
This target atom is the partner in the next collision.

The impact parameter $b$ is given as
\begin{equation}
b = \left|(\mbox{\boldmath $R$}_{\rm A} - \mbox{\boldmath $R$}) \times \mbox{\boldmath $e$}_{\rm p} \right|,
\label{eq:impact_parameter}
\end{equation}
where $\mbox{\boldmath $R$}_{\rm A}$ is the position of the target atom.
Figure 2 depicts the situation described above.
We can obtain the trajectory of particles by the eqs. \ref{eq:scattering_angle}-\ref{eq:impact_parameter}.

Three parameters are defined: bulk binding energy $E_{\rm B}$, displacement energy $E_{\rm d}$, and minimum energy $E_{\rm c}$.
Let us consider a collision from which the original projectile emerges with kinetic energy $E_1$ after transferring kinetic energy $T$ to the target (collision pair) atom.
The target atom is displaced when its kinetic energy exceeds the displacement energy $E_{\rm d}$.
If $T > E_{\rm d}$ is satisfied, the target atom is added to the collision cascade with the kinetic energy $E_2 = T - E_{\rm B}$.
Each projectile is traced while its kinetic energy exceeds a given minimum energy $E_{\rm c}$.

We have extended the ACAT code described above as follows:
While a target atom is produced randomly within a cubic collision cell in the original code, our extended version employs rectangular parallelepiped cells and positions of target atoms are given as initial condition.
The positions of projectile and target atoms are stored in computer memory, which enables us to treat structured target materials as well as amorphous ones.
In the present version it is assumed that each cell contains zero or one target atom.
Thus we can obtain the position of target (collision pair) atom from the stored data when the collision cell is specified.
We call the extended version  "AC$\forall$T" code, which stands for \underline{a}tomic \underline{c}ollisions in \underline{a}ny structured \underline{t}arget.
The notation $\forall$ implies that the code can handle any structure involving crystalline and amorphous.

If the kinetic energy of each projectile $T$ becomes less than a predetermined minimum energy $E_{\rm c}$, for example 200 eV, tracing of the projectile is stopped and its location and velocity are stored, which can be used for initial condition of a succeeding molecular dynamics simulation.

\section{Application of AC$\forall$T code to hydrogen injection into graphite}

A graphite with the size of 321\AA\ (W) $\times$ 347\AA\ (D) $\times$ 335\AA\ (H) is setup as a target material, which consists of 100 layer graphenes.
This material is put into a simulation domain, which is divided into cells whose size is $(\sqrt{3}r_0/2) \times (3r_0/2) \times (h_0)$, as shown in Fig. 3.
Here $r_0 = 1.4463$\AA\ is the distance between neighboring carbon atoms and $h_0 = 3.3480$\AA\ is the distance between graphene layers.
Then each cell contains one carbon atom and the simulation domain includes 256 $\times$ 160 $\times$ 100 cells.

As an example problem, we inject a hydrogen atom with 1.0 keV into the target graphite.
Incident angle is fixed to 0 degree, that is, normal to the target surface.
Incident point is varied for each trial.

Figures 4(a) and 4(b) show trajectories of two cases of an example problem.
Differences of input parameter between two cases are only on the hydrogen incident point.
Black solid circles and gray ones denotes initial points and final points, respectively.

The top black point corresponds to the hydrogen incident point.
The hydrogen atom (projectile) moves along the curve with small angle scatterings.
In a small angle scattering, energy transferred from the hydrogen atom to a target carbon atom is so low that the target cannot be displaced.
When the transferred energy exceeds the displacement energy $E_{\rm d}$, a recoil carbon atom, which initially exists at a black solid circle in the figures, is also traced.
If the kinetic energy of an atom becomes less than the given minimum energy $E_{\rm c}$, tracing the particle is stopped.
The final position is depicted as gray solid circles in the figures.

In figures 4(a) and 4(b), we can see some clusters where a few solid circles gather.
In each cluster, a recoil carbon atom collides with another carbon atom, transfers its own kinetic energy to the collision pair, and the collision pair recoils.
That is, collision cascades occur in each cluster.

In figure 4(b), the curve of trajectory in upper half is almost straight.
In this case the projectile of hydrogen moves straight about 75\AA\ along the interstitial, which may be regarded as a kind of channeling.
This kind of trajectory is not realized in the original ACAT code because it assumes amorphous target.
In figure 4(a), the projectile trajectory just before its stopping lies horizontally with fluctuations.
In this region the projectile travels between graphene layers.
This kind of trajectory is also newly reproduced by our extension of the ACAT code.
We expect that the AC$\forall$T code is useful for evaluation of channeling effects.

\section{Conclusion}
We have extended the ACAT code, which stands for \underline{a}tomic \underline{c}ollisions in an \underline{a}morphous \underline{t}arget and is based on a binary collision approximation, to handle 
any structure involving crystalline and amorphous.
It is shown that new types of trajectories can be reproduced by use of the extended version, named "AC$\forall$T" code.
The name "AC$\forall$T" stands for \underline{a}tomic \underline{c}ollisions in \underline{a}ny structured  \underline{t}arget, and the notation $\forall$ implies that the code can handle any structure involving crystalline and amorphous.

Now we develops a hybrid code of our molecular dynamics (MD) simulation code and binary-collision-approximation (BCA) based code "AC$\forall$T" using a multiple-program multiple-data (MPMD) approach\cite{MPMD}.
As the BCA-based code is relatively lightweight, the resultant hybrid simulation code is expected to simulate carbon materials of submicron order.

The AC$\forall$T code itself holds new features and is promising for evaluation of channeling effects.

\section*{Acknowledgment}
This work is supported by the National Institutes for Fusion Science Collaborative Research Program NIFS09KTAL025 and NIFS10KTAS005, and a Grant-in-Aid for Scientific Research (No. 19055005) from the Ministry of Education, Culture, Sports, Science and Technology, Japan.

\begin{figure}[p]
  \centering
  \includegraphics[width=8cm]{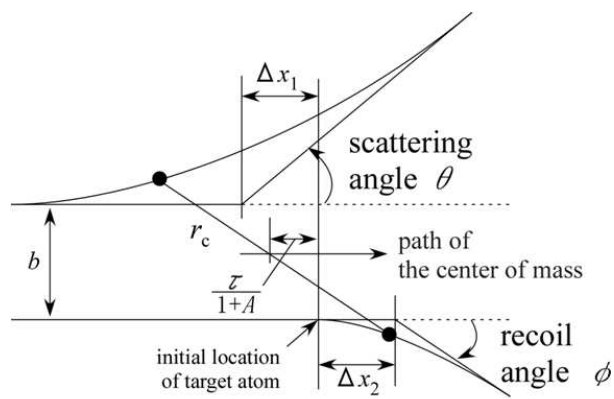}
  \caption{
The trajectory of two particles interacting according to a conservative central repulsive force in the laboratory system (L-system).
The positions of the projectile and the target atom correspond to the apsis of the collision\cite{ACAT}.
  }
  \label{fig:1}
\end{figure}

\clearpage

\begin{figure}[p]
  \centering
  \includegraphics[width=8cm]{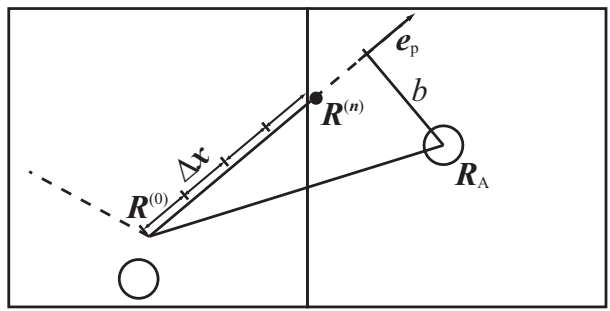}
  \caption{
The procedure of searching the collision partner\cite{ACAT}:
A unit vector $\mbox{\boldmath $e$}_{\rm p}$ is in the direction of a moving projectile.
Projectile starts from $\mbox{\boldmath$R$}^{(0)}$, moves straight in $\mbox{\boldmath $e$}_{\rm p}$-direction with the step length $\Delta x$.
If the unit cell involving the projectile after $n$-th step $\mbox{\boldmath$R$}^{(n)}$ is different from that of the initial position $\mbox{\boldmath$R$}^{(0)}$, a target atom $\mbox{\boldmath$R$}_{\rm A}$ is randomly produced in the new unit cell.
The variable $b$ denotes the impact parameter.
  }
  \label{fig:2}
\end{figure}

\clearpage

\begin{figure}[p]
  \centering
  \includegraphics[width=8cm]{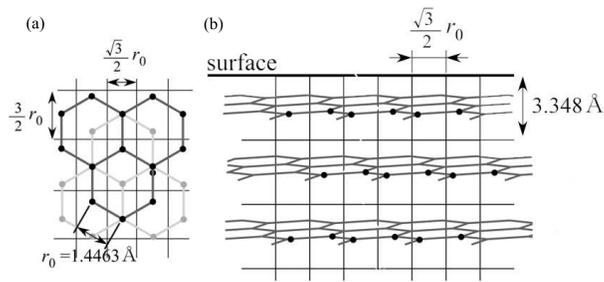}
  \caption{
The configuration of the target materials of graphite and the construction of simulation cell in (a) cross-sectional view, (b) vertical cross-sectional view.
Dots and their linkage denote carbon atoms and bonds.
$r_0 = 1.4463$\AA\ and $3.3480$\AA\ correspond to the length of the bond and the distance between graphene layers, respectively.
In fig. (a), the bonds in black and ones in gray belong to another graphene layer.
Rectangles denote cells for AC$\forall$T simulation.
  }
  \label{fig:3}
\end{figure}

\clearpage

\begin{figure}[p]
  \centering
  \includegraphics[width=8cm]{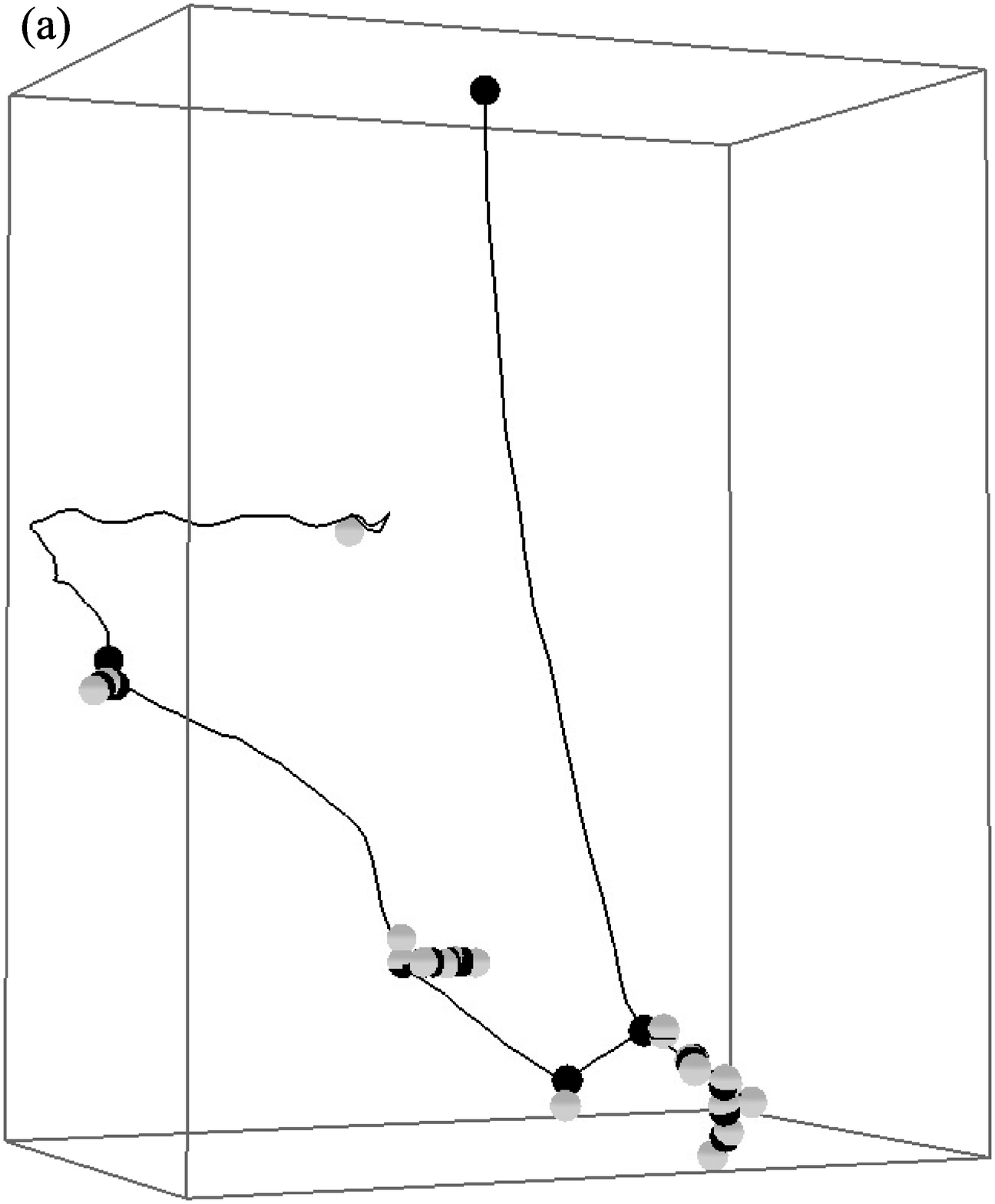}
  \includegraphics[width=8cm]{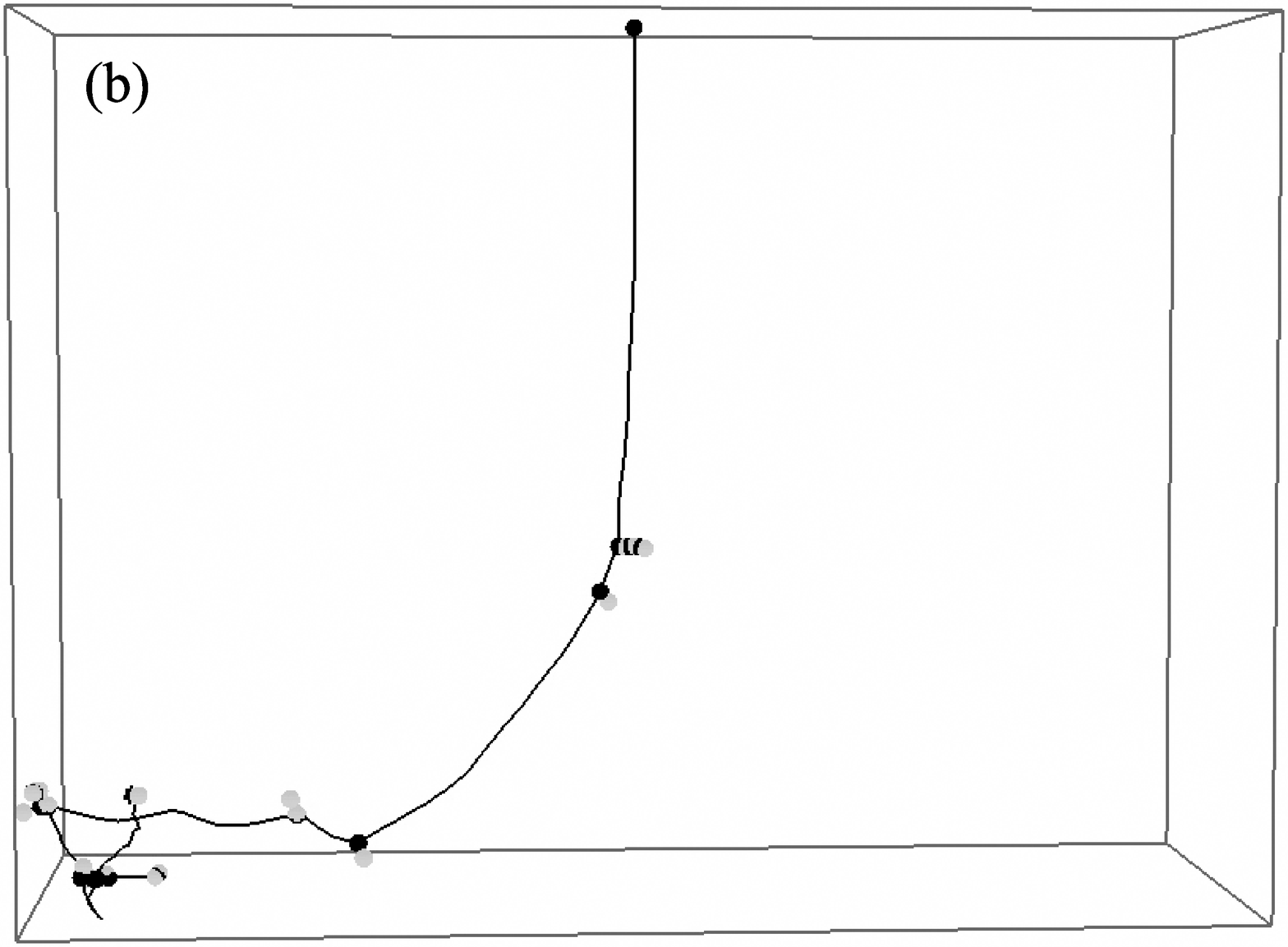}
  \caption{
Trajectory of hydrogen injected into graphite and trajectory of recoil carbon atoms.
Injection energy of hydrogen is 1.0 keV and incident angle is 0 degree.
Incident point differs in (a) and (b).
The size of boxes are (a) 100\AA\ $\times$ 60\AA\ $\times$ 140\AA\ and (b) 250\AA\ $\times$ 120\AA $\times$ 200\AA.
Black solid circles and gray ones denote initial points and final points of the injected hydrogen atom and recoil carbon atoms.
  }
  \label{fig:4}
\end{figure}


\end{document}